\documentclass[a4,useAMS,usenatbib,usegraphicx]{mn2e}
\usepackage{subfig}
\usepackage{lscape}
\usepackage{natbib}
\usepackage{amssymb,amsmath}
\usepackage{fixltx2e}
\usepackage{longtable}
\usepackage[T1]{fontenc}
\title[The Arecibo Galaxy Environment Survey VIII : Discovery of a Nearby Galaxy]{The Arecibo Galaxy Environment Survey VIII : Discovery of an Isolated Dwarf Galaxy in the Local Volume}
\author[R. Taylor, R. F. Minchin,  H. Herbst., R. Smith]{R. Taylor$^{1}$,$^{2}$\thanks{Email: rhyst@naic.edu}, R. F. Minchin$^2$, H. Herbst$^3$, R. Smith $^4$\\
$^1$Astronomical Institute of the ASCR, Bo\v cn\'i II 1401, 14100, Prague, Czech Republic\\
$^2$Arecibo Observatory, HC03 Box 53995, Arecibo, Puerto Rico 00612\\
$^3$Department of Astronomy, College of Liberal Arts and Sciences, University of Florida, 211 Bryant Space Science Center, Gainesville,\\ FL 32611-2055, USA\\
$^4$Department of Astronomy, Esteban Iturra Avenue, Universidad de Concepci\'{o}n, Casilla (P.O.BOX) 160-C, Concepci\'{o}n, Chile.\\
}

\begin{document}

\date{February 2014}

\pagerange{\pageref{firstpage}--\pageref{lastpage}} \pubyear{2013}

\maketitle

\label{firstpage}

\begin{abstract}
The Arecibo Galaxy Environment Survey (AGES) has detected a nearby H\textsc{i} source at a heliocentric velocity of +363 km\,s$^{-1}$. The object was detected through its neutral hydrogen emission and has an obvious possible optical counterpart in Sloan Digital Sky Survey (SDSS) data (though it does not have an optical redshift measurement). We discuss three possible scenarios for the object : 1) It is within the Local Group, in which case its H\textsc{i} properties are comparable with recently discovered ultra-compact high velocity clouds; 2) It is just behind the Local Group, in which case its optical characteristics are similar to the newly discovered Leo P galaxy; 3) It is a blue compact dwarf galaxy within the local volume but not associated with the Local Group. We find the third possibility to be the most likely, based on distance estimates from the Tully-Fisher relation and its velocity relative to the Local Group.

\end{abstract}

\begin{keywords}
galaxies: evolution - surveys: galaxies.
\end{keywords}

\section{Introduction}
The Arecibo Galaxy Environment Survey is an ongoing, fully-sampled neutral hydrogen survey using the ALFA receiver on the Arecibo radio telescope. The survey samples galaxy environments from the Local Void to rich clusters, to study the effects of environment on the H\textsc{i} content of galaxies and resulting effects on their evolution. When complete, it will cover 16 selected areas over a total of 200 square degrees. The survey is bandpass limited to a heliocentric velocity of 20,000 km\,s$^{-1}$ (or 45,000 km\,s$^{-1}$ with new instrumentation). It is a serendipitous discovery within this large, blind volume which we present here.

One of the goals of AGES is to understand the ``missing satellite'' problem. There has long been a discrepancy between the number of galaxies detected in the Local Group and the number predicted in cold dark matter (CDM) simulations. \cite{moore} found that such simulations easily reproduced the correct number of low-mass galaxies in a cluster environment, but failed on galactic scales (with the simulations predicting about a factor 10 more dwarf galaxies than were actually observed).

The discovery of several optically faint galaxies over the last decade presents one possible solution to this problem (\citealt{mcch}) though it is still unclear whether these are sufficiently numerous to fully account for the discrepancy (\citealt{simgeha}) or even whether they are primordial objects at all (\citealt{kroupa}). Finding new satellite galaxies in the Local Group would therefore still have very important implications for understanding galaxy formation.

Blind searches at H\textsc{i} wavelengths do not suffer the same biases as optical surveys, and gas-rich galaxies can be more readily detectable by their H\textsc{i} than optical emission. Two recent discoveries demonstrate the potential of H\textsc{i} surveys for detecting optically faint, nearby dwarf galaxies. ALFA ZOA J1952+1428 (\citealt{travis}) - hereafter AZ1 - is a nearby (7 Mpc) blue compact dwarf galaxy (BCD) discovered by the Arecibo Zone of Avoidance Survey, where H\textsc{i} observations do not suffer from the same extinction effects as optical wavelengths. Leo P (\citealt{leop}) is a galaxy at a distance of 1.7 Mpc (placing it just outside the Local Group, taking the radial extent as 1 Mpc from \citealt{mcch}) discovered by the large-area ALFAFA survey. Though previously catalogued in a search of optical data as a compact group of galaxies by \cite{mcch}, the H\textsc{i} observations revealed that it is in fact a single, very nearby object.

\section{Observations and Results}

We have discovered a hitherto uncatalogued galaxy at a heliocentric velocity of 363 km\,s$^{-1}$. This was found via its H\textsc{i} emission in an AGES survey field targeting the NGC 7448 galaxy group. We have fully described our observations, data reduction and analysis techniques in \cite{me} (hereafter paper VII). In brief, we made a fully-sampled H\textsc{i} survey of a 20 square degree region of the NGC 7448 galaxy group, with a bandpass limit of 20,000 km\,s$^{-1}$ heliocentric velocity ($cz$ = 0.07). We searched the data cube using a combination of visual and automated techniques, our intention being to recover as many detectable H\textsc{i} sources in the cube as possible (regardless of their association with the NGC 7448 target group). We use this sample as a comparison (see section \ref{sec:class}). 

\begin{figure}
\begin{center}
\includegraphics[width=64mm]{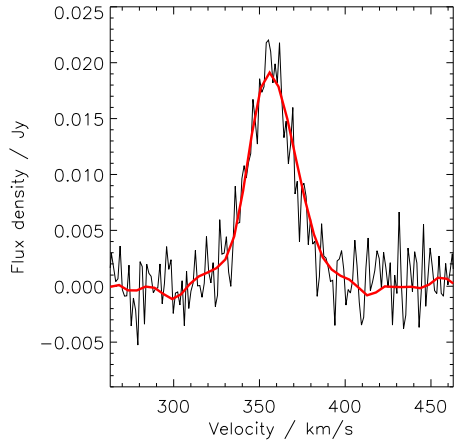}
\caption[spectra]{H\textsc{i} spectra for AF7448\_001. The thick red line is the spectrum from the AGES drift-scan data at 10 km\,s$^{-1}$ resolution; the thin black line is the L-wide follow-up observation at 1.3 km\,s$^{-1}$ resolution.}
\label{fig:hispec}
\end{center}
\end{figure}

\begin{figure}
\begin{center}
\includegraphics[width=64mm]{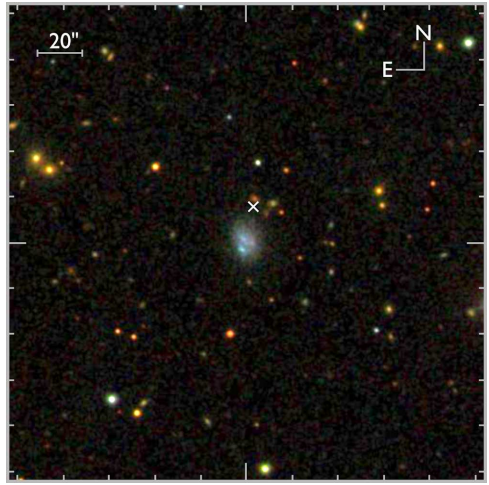}
\caption[rgb]{SDSS RGB image centred on the probable optical counterpart of AF7448\_001. The image spans 3.5$'$. The X indicates the H\textsc{i} coordinates.}
\label{fig:rgb}
\end{center}
\end{figure}

With a peak signal-to-noise ratio (SNR) of 46.0, AF7448\_001 (the name is derived from the survey field, search technique and discovery order) is readily detectable in an H\textsc{i} survey (see figure \ref{fig:hispec} for its H\textsc{i} spectrum). There is only a single plausible optical counterpart visible in the SDSS image at this position (see figure \ref{fig:rgb}). With an offset from the H\textsc{i} coordinates of just 14$''$, this is almost certainly the optical counterpart of the H\textsc{i} detection. Unfortunately there is no optical redshift measurement available - within a 1$'$ radius, no galaxies are listed in NED (though two GALEX UV-sources are identified). While some other small objects are visible close to AF7448\_001, similar objects are found throughout SDSS images. We consider them most likely background galaxies.

The reason this object was not previously catalogued is almost certainly because its apparent optical $g$-band magnitude of 17.06 makes it difficult to detect in an optical survey, as does its small angular size ($D_{\mathrm{opt}}$ = 18.0$''$ at the 26.5 mag arcsec$^{-2}$ isophote). A summary of its properties is shown in table \ref{f01}. It has some similarities to AZ1 - a narrow velocity width, high H\textsc{i} SNR, and an extremely low $M$H\textsc{i}/$L$$_{\mathrm{g}}$ ratio given its optical luminosity (with the caveat of significant distance uncertainty, which we discuss below).

\begin{table}
\small
\begin{center}
\caption[f01]{Observed properties of AF7448\_001. Optical magnitudes and colours have been corrected for external extinction using the model of \cite{schl}. $D$ is the distance to the object in Mpc.}
\label{f01}
\begin{tabular}{l l}\\
\hline
  \multicolumn{1}{c}{Parameter} &
  \multicolumn{1}{c}{Value (error)} \\
\hline
  Right Ascension (J2000) & 22:59:35.3 (10$''$)\\
  Declination (J2000) & 16:46:11 (10$''$)\\
  Galactic longitude & 87.98 \\
  Galactic latitude & -38.40 \\
  $v$$_{\mathrm{hel}}$ (km\,s$^{-1}$) & 363 (2)\\
  W50 (km\,s$^{-1}$) & 23 (3) \\
  $F$$_{\mathrm{H\textsc{i}}}$ (Jy km\,s$^{-1}$) & 0.600 (0.072)\\
  m$_{\mathrm{g}}$ & 17.06 (0.04)\\
  $g$ - $i$ & 0.47 (0.05)\\
  $M$$_{\mathrm{H\textsc{i}}}$/$L$$_{\mathrm{g}}$ & 0.7 (0.1) \\ 
  $M$H\textsc{i} ($M_{\rm \odot}$) & 10$^{5.15}$D$^{2}$\\ 
  $R$$_{\mathrm{opt}}$ (arcsec) & 9.0 (0.4)\\
  $R$$_{\mathrm{opt}}$ (kpc) & 0.045 D\\
\hline
\end{tabular}
\end{center}
\end{table}

AGES has a native resolution velocity of 5 km\,s$^{-1}$, usually Hanning smoothed to 10 km\,s$^{-1}$ to avoid the effects of Gibbs ringing. The narrow velocity width of the source from these data (W50 = 33 km\,s$^{-1}$, W20 = 50 km\,s$^{-1}$) means that instrumental broadening could be significant. We therefore obtained a 5 minute on-off pointing observation using the L-wide receiver at Arecibo, giving a much better velocity resolution of 1.3 km\,s$^{-1}$. Using these data, we found a velocity width of W50 = 23 km\,s$^{-1}$ (W20 = 36 km\,s$^{-1}$), and use these measurements throughout the analysis.

\section{Distance}
A velocity of 363 km\,s$^{-1}$ is equivalent to a distance of 5.1 Mpc if the object is in Hubble flow (with H$_{0}$ = 71 km\,s$^{-1}$Mpc$^{-1}$). In principle, the object is consistent with being in the Local Group given the peculiar velocities of many of the group members (see \citealt{mcch}). However, establishing the true distance to AF7448\_001 is fraught with difficulty.

\subsection{The Tully-Fisher relation}
We can establish a redshift-independent distance estimate using the Tully-Fisher relation (TFR). Fitting our estimates of the W50 (after correcting for inclination, measured using the \textsc{iraf} \textit{ellipse} task) and extinction-corrected $I$-band magnitude (using the approximation of $I$ = $i$ - 0.75, \citealt{win}) to the Cepheid-calibrated relation described in \cite{mast}, we find a distance of 7.9 Mpc. The inclination angle of the galaxy is quite low at 33$\pm$5$^\circ$, making the true velocity width uncertain, so we must treat our estimated velocity width as a lower limit. 

If we make no correction for inclination, we can use the TFR to place a lower limit on the distance estimate. We correct for turbulence using $\Delta t$ = 6.5 km\,s$^{-1}$ as given in \cite{mast}. Using the prescription of \cite{spring}, we make a correction for instrumental effects of $\Delta s$ = 1.0 km\,s$^{-1}$. Following those authors and to minimize the distance estimate, we apply the correction linearly. The spectral broadening due to redshift is negligible. These corrections give an absolute lower limit of the velocity width of 15.7 km\,s$^{-1}$, giving a corresponding Tully-Fisher distance estimate of 1.7 Mpc. A major caveat is that dwarf galaxies do not usually have the flat rotation curves of spiral galaxies which the TFR is calibrated to.

\subsection{Surface brightness - luminosity relation}
\cite{shar} describe a correlation between surface brightness and absolute $V$ luminosity. We convert our photometric measurements into $V$ magnitudes using the prescription of \cite{win}. We find a distance of 3.6 $\pm$ 1.7 Mpc, assuming a distance modulus rms scatter of 1 magnitude (from \citealt{shar}). Since the scatter is influenced by intrinsic variations as well as measurement uncertainties, the error is likely an underestimate.

\subsection{Correction for Local Group velocity}
\begin{figure}
\begin{center}
\includegraphics[width=64mm]{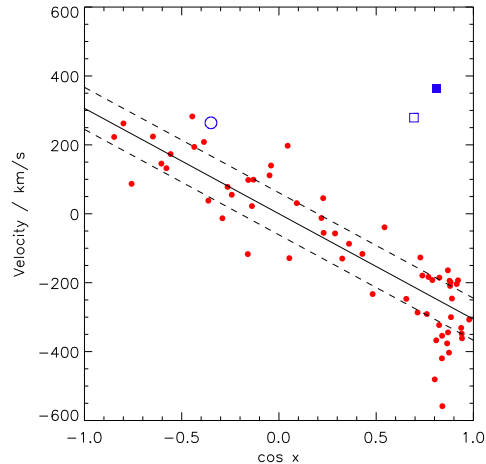}
\caption[gvs]{Heliocentric velocity as a function of cosine of the angular distance from the solar apex, based on the plot of \cite{court} using the data of \cite{mcch} (red circles). AF7448\_001 is shown as a filled blue square while AZ1 is shown as an open square, with Leo P as an open blue circle.}
\label{fig:gvs}
\end{center}
\end{figure}

Another method is to correct for the motion of the object relative to the local group. Figure \ref{fig:gvs} plots the velocities of Local Group members against angular distance from the solar apex. Both AZ1 and AF7448\_001 are moving at completely different velocities to other galaxies at similar angles from the solar apex.

A caveat is that at $cos(\theta)$ $\approx$ 1.0, there are many other galaxies which also deviate from the general trend. These are satellites of Andromeda with strong peculiar motions. While AZ1 is over 67$^\circ$ from M31, AF7448\_001 and M31 are separated by only 33$^\circ$ - not much further than the highest separation of Andromeda's known satellite galaxies (Andromeda XXVIII at 27$^\circ$, or possibly the recently discovered Perseus I at 28$^\circ$ (\citealt{martin}). If AF7448\_001 is a satellite of M31, it would be exceptionally widely separated both in its sky position and velocity. M31 and its satellites are all blueshifted, so the positive heliocentric velocity of AF7448\_001 would be inconsistent with the M31 group dynamics.

The solar motion solution of \cite{court} predicts a velocity of -248 km\,s$^{-1}$ for a galaxy at AF7448\_001's angle from the solar apex, so its velocity relative to the Local Group is 611 km\,s$^{-1}$. This implies a distance of 8.6 $\pm$ 1.7 Mpc (using the rms scatter in the observed local galaxies to estimate the error), broadly consistent with the distance estimated by the TFR.

\section{Classification}
\label{sec:class}
The nature of AF7448\_001 depends upon its distance and whether or not we consider the optical and H\textsc{i} emission to be associated. If they are from separate objects, then the H\textsc{i} properties of AF7448\_001 would be consistent with those of the ultra-compact high velocity clouds (UCHVCs) described in \cite{adams} and \cite{uchvcs}. At 1 Mpc, it would have an H\textsc{i} mass of 1.4$\times$10$^{6}$ $M_{\rm \odot}$ and an H\textsc{i} diameter of $<$ 1 kpc (since it is not resolved by the 3.5$'$ Arecibo beam), somewhat low but consistent with other objects in the \cite{adams} catalogue. Its dynamical mass would be 5$\times$10$^{6}$ $M_{\rm \odot}$, at the low end of the \cite{adams} detections. \cite{adams} propose that such objects may explain the missing satellite problem. However, it seems very unlikely that the H\textsc{i} and optical components of AF7448\_001 are so closely aligned but actually unrelated, especially given the object's isolation (see below).
 
If the optical and H\textsc{i} emission are from the same object, then it is much less likely that AF7448\_001 resides in the Local Group. The lowest possible distance would be about 1.7 Mpc from the TFR, in which case the object is broadly similar to Leo P (\citealt{leop}). It would have about half the H\textsc{i} and dynamical mass of Leo P (assuming that $R_{\mathrm{H\textsc{i}}}$~=~1.7$R_{opt}$ from \citealt{swat}). Its stellar mass, using the parameters of \cite{bell}, would be 5.1$\times$10$^{5}$ $L_{\rm \odot}$ (based on its $g$ - $i$ colour), only slightly higher than the 3.6$\times$10$^{5}$ $L_{\rm \odot}$ of Leo P.

\begin{figure}
\begin{center}
\includegraphics[width=64mm]{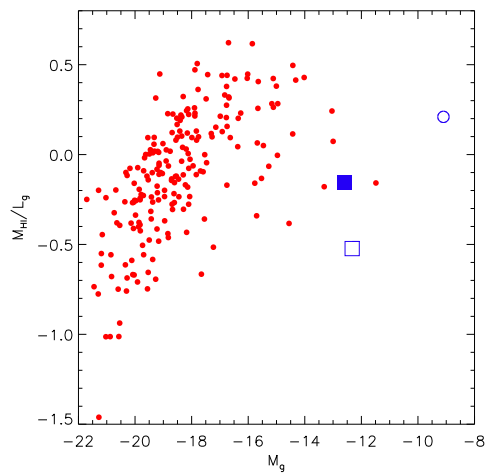}
\caption[gvs]{$M$H\textsc{i}/$L_{\mathrm{g}}$ ratio (logarithmic scale) as a function of M$_{\mathrm{g}}$. AF7448\_001 is shown as a large filled blue square while the small red squares, for comparison, are the entire sample for the data set in which it was detected. AZ1 is shown by an open blue square while Leo P is an open blue circle.}
\label{fig:MHIL}
\end{center}
\end{figure}

\begin{figure}
\begin{center}
\includegraphics[width=64mm]{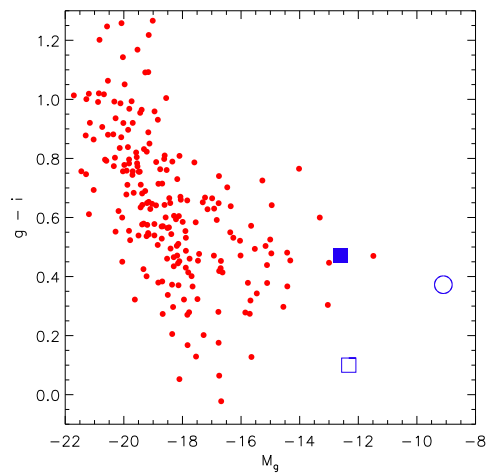}
\caption[gvs]{Colour-magnitude diagram with AF7448\_001 shown by a filled blue square. The red squares show the other H\textsc{i} detections in the data cube in which it was detected. AZ1 is shown by an open blue square while Leo P is an open blue circle.}
\label{fig:CMD}
\end{center}
\end{figure}

Being in or just behind the Local Group requires an unusually high peculiar velocity for AF7448\_001. The various distance estimates suggest it is actually a few Mpc away. At 8.6 Mpc, AF7448\_001 has an H\textsc{i} mass of 1$\times$10$^{7}$ $M_{\rm \odot}$, an absolute $g$ magnitude of -12.61 (equivalent to 1.5$\times$10$^{7}$ $L_{\rm \odot}$) and an optical diameter of approximately 0.8 kpc. Its $M$H\textsc{i}/$L$$_{\mathrm{g}}$ ratio of 0.7 is extremely low given this luminosity (see figure \ref{fig:MHIL}); while its $g-i$ colour of 0.47 is somewhat red it could not be described as a red sequence object (see figure \ref{fig:CMD}). Morphologically it is not smooth and shows several bluer substructures which may be star-forming regions (see figure \ref{fig:rgb}). Overall, its properties are consistent with being a BCD (see \citealt{vanzee} and \citealt{hucht} for comparison).

The very low $M$H\textsc{i}/$L$$_{\mathrm{g}}$ ratio and slightly red colour raises the question of whether AF7448\_001 is H\textsc{i} deficient. Using the parameters of \cite{gv} for BCD galaxies gives a deficiency of -0.2, hence the object is not deficient. The $M$H\textsc{i}/$L$$_{\mathrm{g}}$ ratio is relatively low, but there is a selection bias in that galaxies with higher $M$H\textsc{i}/$L$$_{\mathrm{g}}$ ratios are easier to detect, especially at these very low luminosities (which we can only detect at all in the very nearby Universe).

\begin{figure}
\begin{center}
\includegraphics[width=70mm]{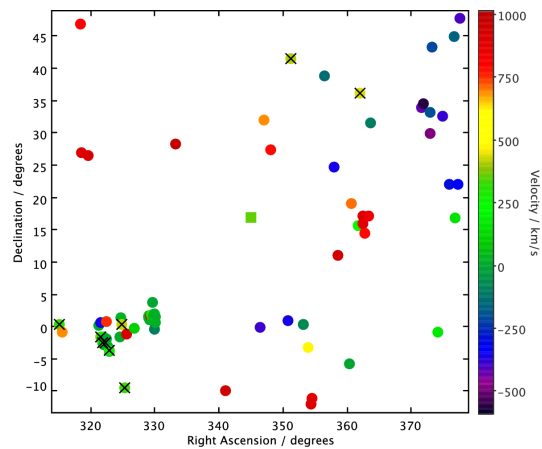}
\caption[gvs]{Galaxies from a NED search within a 33$^\circ$ radius of AF7448\_001 (central square; other galaxies are shown as circles). Galaxies with a velocity $\pm$ 100 km\,s$^{-1}$ of AF7448\_001 are highlighted with crosses. Velocities are heliocentric.}
\label{fig:isolation}
\end{center}
\end{figure}

AF7448\_001 appears to be an isolated galaxy, with no other galaxies detected by AGES (or any bright galaxies listed in NED) in the 20 square degree survey region with $cz$ $<$ 1,500 km\,s$^{-1}$. A NED query within a 33 degree radius (5.0 Mpc at 8.6 Mpc distance) finds that the nearest bright galaxy at a similar ($\pm$ 100 km\,s$^{-1}$) velocity is NGC 7640, which is 24.6$^\circ$ degrees away (3.7 Mpc). AF7448\_001 easily passes the isolation criteria of \cite{karch}, with no (known) galaxies of similar optical diameter within a projected distance of 20 optical diameters (see figure \ref{fig:isolation}). Its isolation is comparable to the recently-discovered GHOSTS~I described in \cite{monach}.

\section{Summary}
We have discovered a nearby galaxy at heliocentric velocity of 363 km\,s$^{-1}$. This object has evaded all previous optical catalogues, but is readily detected in the H\textsc{i} line. The galaxy is most likely a BCD and is broadly similar to another object discovered by a blind H\textsc{i} survey, AZ1.

Of the three distance possibilities we considered - within the Local Group, at the Local Group outskirts, or in the nearby volume - the latter seems the most likely. The first two would require a much higher peculiar velocity than any known Local Group members. It is more likely that this is a nearby, isolated BCD at a distance of a few Mpc.

The apparent isolation of AF7448\_001 makes it a worthwhile target for future, higher spatial resolution observations. Using resolved H\textsc{i} maps from the Australia Telescope Compact Array, \cite{ls} find evidence that even apparently isolated BCDs are interacting and therefore not really isolated. In contrast, \cite{uson} find that the BCD NGC 2537 is likely evolving in isolation (despite its close proximity on the sky - 16.7$'$ - to IC 2233), with no signs of disturbance in its H\textsc{i} map from the Very Large Array. The blue colour of BCDs has been variously attributed to starbursts (e.g. \citealt{zhao}) and low metalicities (e.g. \citealt{leop}, \citealt{bellz}). Optical spectroscopy would potentially confirm the purported optical counterpart of AF7448\_001, and could also address whether it is a relatively primordial object or a more evolved galaxy currently experiencing a high star formation rate.

We note that while AF7448\_001 is almost certainly not a Local Group object, its discovery raises the question of how many similar, perhaps closer, objects could have escaped cataloguing in the SDSS data. The completion of larger area H\textsc{i} surveys, such as ALFALFA and ALFA ZOA, may yet provide solutions to the missing satellite problem.

\section*{Acknowledgments}

This work was supported by the project RVO:67985815. R.S. acknowledges support by FONDECYT grant 3120135.

This work is based on observations collected at Arecibo Observatory. The Arecibo Observatory is operated by SRI International under a cooperative agreement with the National Science Foundation (AST-1100968), and in alliance with Ana G. M\'{e}ndez-Universidad Metropolitana, and the Universities Space Research Association. 

This research has made use of the NASA/IPAC Extragalactic Database (NED) which is operated by the Jet Propulsion Laboratory, California Institute of Technology, under contract with the National Aeronautics and Space Administration. 

This work has made use of the SDSS. Funding for the SDSS and SDSS-II has been provided by the Alfred P. Sloan Foundation, the Participating Institutions, the National Science Foundation, the U.S. Department of Energy, the National Aeronautics and Space Administration, the Japanese Monbukagakusho, the Max Planck Society, and the Higher Education Funding Council for England. The SDSS Web Site is http://www.sdss.org/.

The SDSS is managed by the Astrophysical Research Consortium for the Participating Institutions. The Participating Institutions are the American Museum of Natural History, Astrophysical Institute Potsdam, University of Basel, University of Cambridge, Case Western Reserve University, University of Chicago, Drexel University, Fermilab, the Institute for Advanced Study, the Japan Participation Group, Johns Hopkins University, the Joint Institute for Nuclear Astrophysics, the Kavli Institute for Particle Astrophysics and Cosmology, the Korean Scientist Group, the Chinese Academy of Sciences (LAMOST), Los Alamos National Laboratory, the Max-Planck-Institute for Astronomy (MPIA), the Max-Planck-Institute for Astrophysics (MPA), New Mexico State University, Ohio State University, University of Pittsburgh, University of Portsmouth, Princeton University, the United States Naval Observatory, and the University of Washington.

{}

\label{lastpage}

\end{document}